# IoT Applications in Urban Sustainability

Samiya Khan[1,a], Mohammad Moazum Wani[2,b], Mansaf Alam[1,c]

[1]Department of Computer Science, Jamia Millia Islamia, India

[2]Department of Information Technology, Central University of Kashmir, India

[a]samiyashaukat@yahoo.com, [b]moazumwani21@gmail.com, [c]malam2@jmi.ac.in

**Abstract**: *Internet of Things is one of the driving technologies behind the concept of Smart Cities and is capable of playing a significant role in facilitating urban sustainable development. This chapter explores the relationship between three core concepts namely Smart Cities, Internet of Things and Sustainability; thereby identifying the challenges and opportunities that exist in the synergistic use of Internet of Things for sustainability, in the Smart Cities context. Moreover, this chapter also presents some of the existing use cases that apply Internet of Things for urban sustainable development, also presenting the vision for these applications as they continue to evolve in and adapt to the real world scenario. It is because of the interdisciplinary nature of these applications that a clear comprehension of the associated challenges becomes quintessential. Study of challenges and opportunities in this area shall facilitate collaboration between different sectors of urban planning and optimize the utilization of Internet of Things for sustainability.*

## 1   Introduction

Sustainability is an established concept in urban planning. With the advent of the concept of Smart Cities, driven by Internet of Things (IoT), the notion of sustainability has reached another level of maturity. Many of the major concerns of urban sustainability such as waste management and environment preservation can be resolved using IoT inherently for the greater benefit of mankind.

Smart Cities is a popular concept that integrates many technologies to create a comprehensive system. This system envisions improving the quality of life and citizen satisfaction. This aspect of Smart Cities relates it to the objectives of sustainability in urban areas closely. Besides many other technologies such as Cloud Computing and big data analytics, Internet of Things (IoT) forms one of the core technologies that drive Smart Cities and inculcates the required intelligence in its sub-systems. However, the use of IoT for sustainable development of urban cities is not short of challenges. This chapter examines the relationship between IoT and urban sustainability, looking at the hurdles that exist in its pathway and suggests possible solutions.

The rest of the chapter is organized as follows: Section 2 gives a background of Smart Cities, Internet of Things and sustainability, establishing associations between these entities. Section 3 elaborates on some of the widely accepted and popular use cases of IoT for sustainability. Section 4 investigates the challenges that persist in the use of IoT for sustainability and provides insights on possible solutions. Lastly, Section 5 synopsizes the chapter, throwing light on scope for future research in this area.

## 2   Background

This section provides a background of technological paradigms such as Internet of Things (IoT) and Smart Cities. In addition, it elaborates on the concept of sustainability, also examining the associations that exist between these three conceptual entities.

### *2.1 Smart Cities and Internet of Things*

The concept of Smart Cities has gained immense popularity in the recent past because of its ability to use technological innovations for the benefit of citizens and humanity, in general.

According to a study (Bakıcı, Almirall, & Wareham, 2013), about 50% of the total world population lives in urban areas. With that said, as the urban population has increased, the services provided to them have deteriorated both in terms of quantity as well as quality.

The smart city paradigm envisions providing high-quality services to citizens. Moreover, Government and private institutions back the use of ICT technology for improving the operational efficiency of the system by devising sustainable solutions to challenges that the society faces today (Su, Jie, & Hongbo, 2011, Caragliu, Del Bo, & Nijkamp, 2011). Some of the sectors that are known to have been at the center of attention include energy, crime management, heath, education, waste management, traffic and unemployment (Chourabi et al., 2012).

Existing literature indicates disparate viewpoints on the way in which intelligence must be infused to transform a city into a smart city. Most of the available studies support the standpoint that intelligence must be infused into the sub-systems such as education, energy infrastructure, waste management, water system, healthcare and infrastructure, which can then be integrated to create a unified system (Kanter & Litow, 2009, Gurdgiev & Keeling, 2010). Studies on smart city frameworks (Perera, Zaslavsky, Christen, & Georgakopoulos, 2014; Giffinger et al., 2007) have identified six components, which include smart living, smart economy, smart governance, smart people, smart environment and smart mobility. These components are illustrated in Fig. 1.

The advent of Internet of Things has revolutionized smart city construction in more than one way. Until recently, urban information systems were limited to simple querying, without any real analysis being presented. However, with the evolution of urban systems, particularly in the smart city context, data is integrated from multiple sources and analyzed to present temporal analysis, which can significantly facilitate urban management and associated decision-making. Although, research is still underway in the analysis and visualization of real time data, the smart city infrastructure is way ahead of its predecessor.

Recent past has seen an upsurge in sensor deployments because of reduction in production cost of sensors. Besides this, advances in cloud and sensor technology have also contributed to this rise. The evolution of smart city and IoT has taken place because of disparate reasons. Smart Cities have advanced as a result of changing user needs and the demand for novel applications. On the other hand, IoT has progressed innately from the technological advances.

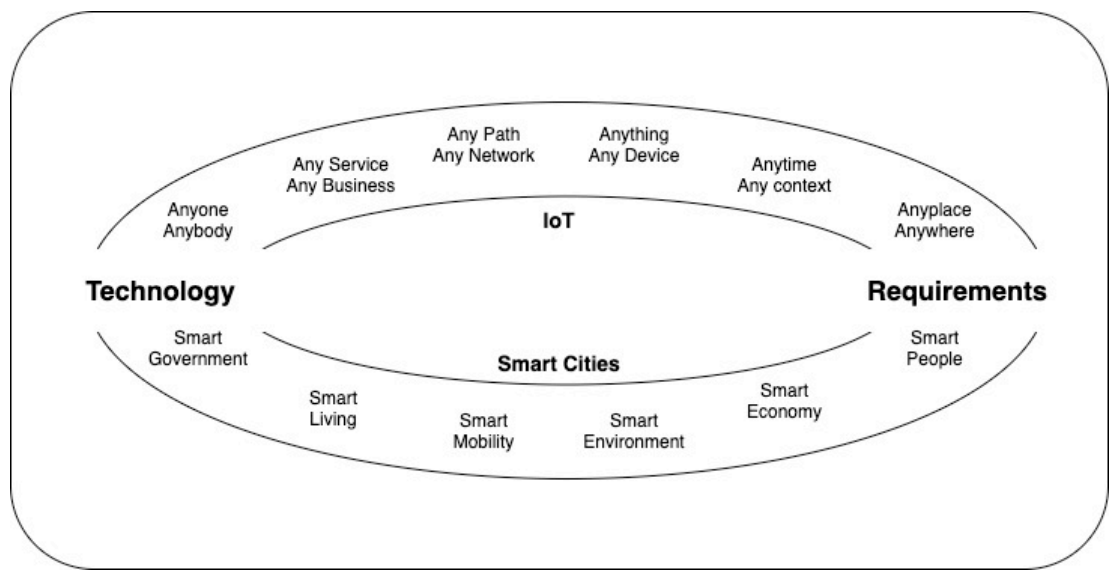

Fig. 1 - Relationship between Smart Cities and IoT (Samih, 2019)

The most widely accepted definition of Internet of Things (IoT) is a system that allows things and people to be connected to each other via the Internet beyond the limits of time and space. Regardless of the differences between these two technological paradigms, they are racing towards each other to accomplish the common objective of benefitting the society. An illustration of how these concepts are related to each other is given in Fig. 1.

## 2.2 Concept of Sustainability

Sustainability is a well-known concept in the field of urban planning. Several definitions and elaborations have been presented in literature. However, there is no standard definition (Gatto, 1995). In view of the fact that this chapter focuses on sustainability from the perspective of how it can be achieved with the use of IoT, we focus on the aspect that describes sustainability as a mode of development that balances demand and supply in a coherent manner. Sustainable planning is expected to provide the following three outcomes (American Planning Association, 2016):

- It should include a plan for ensuring equality.
- Communities created as a result of implementation of sustainability must demonstrate diversity, resilience and self-sufficiency.
- Implementation of sustainable principles must add to creation of a 'healthy' environment from the perspective of natural resources' usage as well as improving social and economic health of the system.

From the American Planning Association's description of sustainability, it can be understood that this concept goes beyond the use of natural resources and products. It must impact and improve multiple facets of the society, both individually and as a whole.

*2.3 Role of IoT in Sustainability*

Internet of Things is one of the driving technologies behind sustainable development. Moreover, the linkage between Internet of Things and Smart Cities is well established. Therefore, a parallel can be drawn between smart city concept and sustainable development. The concept of Smart Cities can be seen as an extension to sustainability initiatives. The fundamental goal of sustainable development is to maximize benefits to the citizens in such a way that costs and societal impact are minimized (American Planning Association, 2016). The concept of Smart Cities is also based on the same principle, supporting and extending the idea of sustainable development.

## 3   IoT Use Cases for Urban Sustainability

It would not be wrong to state that this is the era of Internet of Things. There is an exponential rise in the use of sensors, wearable smart devices and appliances  (Khan & Alam, 2020). This technological innovation has impacted every walk of human life and sustainable development is no exception to this rule. This section presents some use cases of IoT for sustainability. Statistical evidence to support the use of IoT in the application area concerned is provided in Table 1.

Table 1 – Statistics: IoT for Sustainability Use Cases

| S. No. | Use Case | Statistical Data/Trends |
|---|---|---|
| 1. | Smart Grids | 1. Environmental Defense Fund predicts a cut of 30% in air pollution with the use of smart grids (Krupp, 2011). 2. Savings in energy with the use of smart grids is estimated to be equivalent to 70 million road trips around the world (Smart Energy Consumer Collaborative, n.d.). |
| 2. | Smart Meters | 1. Giving real time data to users about their consumptions is estimated to impact 40% of consumption patterns for a building (Marinakis, 2020). |
| 3. | Smart Lighting | 1. Smart LED lights consume 50% of the electricity used by conventional lighting systems (Mishra, 2018). 2. enModus reported that intelligent systems that use smart LED lights can reduce energy consumption by as much as 99% (enModus, 2018). |
| 4. | Smart Streetlights | 1. According to a study by Navigant Research, a further saving of 10-20% is achieved with the use of natural light adjustment and intelligence (Bjorlin, 2017). |
| 5. | Smart Water | 1. Sensus estimated a reduction in leakage waste by 20% after the incorporation of IoT in water management systems (Sensus, n.d.). |
| 6. | Smart Waste Management | 1. Bigbelly reports that collection instances are reduced by 80% with the use of smart waste management systems (BigBelly, |
| | | 2019). |
| 7. | Sensor-based Air Quality Monitoring | 1. The sensors available for air quality monitoring can collect location specific data in real time. The sensitivity of these sensors is as high as 0.1 parts per million for Libelium-detecting sensors (WASP MOTE, 2017). |

### 3.1 Smart Grids

Intelligent systems for distribution of electricity allow service providers to manage their assets in an effective manner. In this age of connected systems, it is possible for companies to reroute and manage distribution even in the case of outages (Yu & Xue, 2016). Prorating of electricity on the basis of demand and supply is also possible, which is particularly relevant for cases where renewable energy sources are involved.

In addition to better management of assets, smart grids also allow providers to detect outages by continuous monitoring of infrastructure. They are no longer dependent on customer complaints for this purpose. From the customer's perspective, smart grids allow better quality of service with improved efficiency and streamlined maintenance. The integration of smart grids with the smart homes infrastructure shall now make it possible to give usage feedback to users. This shall further allow them to optimize their consumption and reduce bills.

### 3.2 Smart Meters

Smart meter, in addition to energy usage data, also collects real time data related to consumption of water and gas (Venables, 2007; Hauber-Davidson, 2006). Therefore, smart meters are a component of the smart grid infrastructure as well. The difference between a smart meter and a conventional meter is that unlike the latter that generates a bill at the end of the month, smart meter provides information about usage in real time. Therefore, users can optimize their consumption habits and monitor the same for their benefits. From the companies' perspective, it is convenient for them to monitor the system for failures and respond to maintenance issues with the incorporation of smart meters.

### 3.3 Smart Lighting

One of the most popular use cases of IoT for sustainability is smart lighting. The use of electricity can be significantly optimized with the use of intelligent systems (Castro et al., 2013). Natural light cycles can be mimicked by incorporation of light and temperature sensors. In view of the fact that a majority of people spends more time indoors than outdoors, this can have a significant impact on energy consumption. Moreover, light sensor-based applications to manage the orientation of solar panels for optimal use of natural resources can also be attempted in the future.

### 3.4 Smart Streetlights

Smart street lighting is a special case of smart lighting where streetlights are connected to an intelligent system to take advantage of natural light and reduce energy consumption whenever and wherever possible. This infrastructure can be integrated with other systems such as air quality monitoring, security infrastructures like cameras and traffic management (Wani et al., 2020) for development of a comprehensive system for infrastructure management of streets (Sudheer et al., 2019).

### 3.5 Smart Water

Water is consumed by disparate infrastructures ranging from residential and commercial complexes to agricultural and industrial settings (Garg et al. 2020). Smart water systems have distinctive uses on the basis of the target infrastructural setting. In agricultural and lawn settings, water irrigation systems are getting smarter for monitoring soil saturation and prevention of under or over watering of the soil (Kamienski et al., 2019). Many other applications also exist in the agriculture sector such as smart irrigation systems. In addition, water sensors are also used to quantify and measure quality and check for presence of chemicals and wastewater in usable water supplies (Prasad et al., 2015).

Smart water system can also be integrated with other systems like weather forecasting system to monitor and control drainage. They may also be integrated with home infrastructures to warn owners against potential leaks and damage, thereby developing a comprehensive water management infrastructure. Besides this, future applications in this area include automatic detection of incoming and terminating water supply for efficient usage of water to prevent wastage.

### 3.6 Smart Waste Management

One of the main facets of environmental sustainability is effective waste management. Smart waste management systems have a connected trash can and waste collecting vehicles. The trash cans inform the nearest trash collection vehicle when it is full and the vehicle is informed about the shortest route to the trash can by an integrated traffic management and forecasting system (Folianto et al., 2015). In this way, a smart waste management system can optimize costs by reducing fuel usage and manages traffic to eliminate scenarios of congestion due to trash collection vehicles. Other such systems can also allow users to locate and identify recycling opportunities, which support sustainability in the truest sense.

### 3.7 Sensor-based Air Quality Monitoring

Sensors are extensively being used for making different parametric assessments of indoor (Patil et al., 2019) and outdoor (Kaivonen & Ngai, 2020) air and its quality. Some of the captured parameters include temperature, carbon dioxide, pressure and humidity levels. Besides this, specialized sensor may be used to detect the existence of organic compounds like ozone, black carbon and methane, in addition to many others. Analysis of this data is further used to detect high-pollution localities and sources of pollution. Consequential decision-making and interventions may then be made to alleviate the issue.

### 3.8 Other Applications

In addition to the above mentioned, there are many other use cases of IoT for sustainability. For instance, Internet of Things is known to be playing an influential role in sustainable product development. Smart design methodology is increasingly being used to optimize resources and their consumption during manufacturing of products (Qu et al., 2019). IoT is also being integrated into the supply chain to improve transparency of the system.

Apart from this, IoT can provide contributory insights into real time detection of disease outbreaks in animals by fitting special RFIDs into animals like livestock (Yang & Johnson, 2020). In this manner, acute scenarios can be avoided, with data being available for designing intervention plans. Besides this, wildfires and deforestation scenarios (Hidestål & Zreik, 2020) can also be monitored using smart systems. The use cases mentioned in this section are just a few of the ways in which IoT can be used for sustainability and is just the tip of the iceberg, with immense scope for future research.

## 4   IoT Challenges and Vision for Sustainability

Undoubtedly, Internet of Things can push and enhance urban sustainability by leaps and bounds. However, this pathway is not short of challenges. This section throws light on the challenges involved in using IoT for urban sustainability (Zhang, 2017) and provides insights on some of the approaches that may be used to mitigate these challenges.

### 4.1 Span

The geographical coverage of sensor devices is quantified in terms of span. Therefore, the term span refers to the density of sensor devices installed in an area and exhibits dependence on three factors namely budget, policy and installation locations. Locations for installation are selected on the basis of application requirements. Therefore, the parametric values that need to be detected form one of the bases in this selection. Besides this, other factors like operational and maintenance requirements of the system are also considered before making this selection. From the perspective of citywide sustainability infrastructure, coverage and installation location selection shall require consent and advice from policy makers, technicians and planners.

Understandably, a higher number of sensors shall result in better precision as the network will become denser and consequently offer improved coverage. However, such an ensemble increases the costs involved in infrastructure development. Designing a sensor network for urban sustainability infrastructure requires technicians. However, the role of planner cannot be undermined because installation of devices may or may not change the existing plan of the city. Decisions like whether sensors can be installed in existing devices or newer devices with embedded sensors need to be implanted require assistance from both technicians and urban planners. Besides this, it can affect the costs of the system as well.

A conceptual association lies between Smart Cities and the concept of sustainability, notwithstanding that IoT is the driving technology behind the former. In order to elaborate on the role of IoT within a sustainable infrastructure, it is important to understand the position of this technology in the process of sustainable development. IoT backs up the phase of data acquisition by allowing the sensor network to capture data required by specific applications such as evaluation of system parameters and predictive or prescriptive analytical solutions for sufficing specific needs of the system. Therefore, IoT is positioned at the preliminary stages of data acquisition within the system, which is a critical stage in view of the fact that compromised data quality will directly result in poor analysis. Although, span quantifies data detection strength, it can also entail data range requirements at the broader planning level.

## 4.2 Fault Tolerance

Fault tolerance is a term used to describe the resilience of a system to failure and its ability to respond to user requests under such circumstances. One of the critical factors associated with fault tolerance in urban sustainability is data security. Sensors have lower information capacities. Moreover, it is usually not possible to incorporate security protection methods in sensor networks when used at the civic level (Atzori et al., 2010). Another aspect of fault tolerance that continues to exist in the sustainable smart city is questionable optimization. This stems from the fact that Smart Cities prioritize user experience.

In order to mitigate the challenges associated with fault tolerance, the notion of resiliency must be incorporated into the quest for sustainability. It is important to understand that in the current context, the standard definition of resilience will also need to evolve from 'ability of a system to go back to constant equilibrium state' (Pimm, 1984) to 'ability of a system to adjust to new equilibrium' (Holling, 1973); emphasizing on adaptability as the most crucial facet for dynamic environments.

## 4.3 Data Ownership

Existing literature suggests that improving operational efficiency is the one of the most widespread uses of IoT for city infrastructures. Since, such applications totally rely on acquired data, ownership of data becomes a major concern. Most of the data used for such applications include user data or data associated with citizens. In order to utilize the full capacity of the system, this data needs to be in the public domain. However, bringing user data to the public domain may breach confidentiality and privacy requirements of such systems. Finding a balance between these two facets is a significant challenge (Janis et al., 2012).

To deal with such challenges, urban planners must make significant efforts for bridging the gap between the system and users. The users must be aware of how their data is being used and for which applications. This is particularly important for applications that make direct use of citizen data coming from smart homes or smart healthcare systems.

### 4.4 Lack of Incentive

It can be seen that the smart city initiative is a collaborative effort between enterprises and Governments (Greenfield, 2013). However, what this also means is that there might be a lack of initiative by the enterprises considering the constrained profit potential of such applications. In order to deal with such issues, other features may be incorporated into sustainable infrastructures, facilitating collaboration between different sectors. For example, real estate developers and technological counterparts can collaborate for smart city laterals like smart homes, catalyzing the use of IoT for sustainability.

### 4.5 Adverse Technological Effects

Infrastructures developed and planned for urban sustainability envisage citizen benefits and put them at highest priority. However, such infrastructures may have potentially adverse effects. Firstly, the construction of sensor networks may involve large energy costs. Moreover, the use of these systems may also result in relatively higher energy usages (Van den Bergh, 2011). Therefore, the outcomes of system adoption may not be as planned by policy makers. Dealing with these issues involves cooperation between policy makers, urban planners and technicians to strategize ways in which this new system can be adopted. Multi-faceted impact analysis must be done in advance to predict possible issues and plan their alleviation. Although, the actual impact can only be determined after the system becomes operational, strategizing and planning may help in curtailing potential concerns.

## 5   Conclusion

In this era of data science and artificial intelligence, the core technology that provides that data required by these technologies to function at their best capacities is Internet of Things. Although, the position of IoT in the data lifecycle lies in data acquisition, it is central in view of the fact that the results and precision of the derived results determined by all the other technologies depend on data availability and quality. Smart Cities integrate multiple technologies to provide smart services to citizens. The primary objective of urban sustainability is to improve citizen services. Thus, Smart Cities and urban sustainability work to a common goal. IoT is being put to use in multiple ways to support sustainability in Smart Cities. However, challenges such as fault tolerance, span, data ownership and institutional issues exist. Possible solutions to these challenges have been proposed in this chapter. Future work in this area shall include implementation of solutions and development of frameworks for evaluation of the cost-precision tradeoff that continues to exist in the use of IoT for sustainability.